\documentstyle[sprocl,epsfig]{article}
\begin{document}
\title{CP Violation and Weak Decays}
\author{R. D. Peccei}
\address{Department of Physics and Astronomy, UCLA, Los Angeles, 
CA 90095-1547}
\maketitle
\begin{abstract}
I review some of the salient issues connected to CP-violation and weak decays.
In particular, I focus on recent and forthcoming tests of the CKM model and discuss the accuracy one can expect to achieve in B-decays for the parameters of this model.

\end{abstract}

\section{Introduction}

There is a large difference in emphasis between the study of weak
decays and decays which lead to CP-violating effects.  The former,
paradoxically, essentially test our understanding of the strong
interactions, while CP-violating phenomena explore the limits
of our knowledge.  Of course, in practice,
both phenomena are linked and the extraction of CP-violating
effects is often clouded by our imperfect understanding of the strong
interactions.  We will see a case in point below when we discuss
$\epsilon^\prime/\epsilon$.

In view of these considerations, it has proven natural to focus
one's attention in systems like B-decays where one can best control
the effects of the strong interactions.  Indeed, matters simplify
considerably if one looks at heavy quark systems where
$m_Q \gg \Lambda_{\rm QCD}$.  This can perhaps best be appreciated
by considering the ratio of charged to neutral lifetimes in heavy-light $q\bar Q$ bound states.  For Kaons, where
$m_Q \equiv m_s \sim \Lambda_{\rm QCD}$, this ratio is nearly 140.
For $D$ mesons, where $m_Q \equiv m_c \sim (5-10)\Lambda_{\rm QCD}$,
this ratio is reduced to about 2.5.
However, for $B$ mesons, where $m_Q \equiv m_b \sim (15-30)\Lambda_{\rm QCD}$, this ratio is very close to unity: $\tau(B^+)/\tau(B^0) = 1.072 \pm 0.026$ .~\cite{PDG}
Although one has some understanding why these $\tau_+/\tau_0$
enhancements occur,
~\cite{deltaI,GNPR} clearly life is simpler for $B$ decays.  Here, to a first approximation, one can neglect the effects of the light-quark
spectator, so that $B$-decays are essentially $b$-decays. 
Furthermore, for $B$-decays one can use a systematic expansion
in the heavy quark mass $m_b$---the, so called, Heavy Quark
Effective Theory (HQET)~\cite{HQET}---to effectively incorporate
corrections of $O(\Lambda_{\rm QCD}/m_b)$ to the simple
spectator approximation.

\section{CP Violation--Preliminaries}

Although the study of CP violation is a mature subject, we
still have very limited experimental information.  This consists of:
\begin{description}
\item{i)} Measurements of certain CP-violation parameters in the
$K^0-\bar K^0$ complex.  Until recently, all these measurements
could be explained in terms of the $\Delta S=2$ complex
parameters $\epsilon$.  However, very recently, new evidence
for a non-vanishing value for the $\Delta S=1$ parameter
$\epsilon^\prime$~\cite{KTeV} has been announced.
\item{ii)} Bounds on other CP-violating or T-violating quantities.
Prototypical of these are the strong bound on the electric dipole
moment of the neutron [$d_n \leq 1.2\times 10^{-25}~
{\rm e~cm}$ (95\% C.L.)~\cite{PDG}].
\item{iii)} Indirect evidence from cosmology.  Here the strongest
evidence is the observed baryon---antibaryon asymmetry of the
Universe, embodied in the ratio $\eta_B = (n_B-n_{\bar B})/n_{\gamma}$ which is of order
$\eta_B \sim (3-4)\times 10^{-10}$.~\cite{asym}
\end{description}
Theoretically, one is not much better off.  Although we understand the
framework needed to have CP-violation in a theory, 
we really still do not understand the details of how CP-violation occurs in nature. In particular, even though we have a working paradigm for CP-violation, the CKM~\cite{CKM} model, we actually have no real
proof of the validity of this paradigm.  In fact, we have indirect evidence that the CKM paradigm must fail at some level!

It is useful to describe the present theoretical prejudices regarding CP-violation.  The first of these is that the observed
CP-violation is due to an explicit breaking of CP at the Lagrangian level,
and not as a result of spontaneous CP-violation.  This latter possibility is
disfavored by cosmology, as it leads to too much energy density in the domain walls separating different CP-domains in the Universe.~\cite{KOZ} The second
prejudice is that CP violation is connected with renormalizable interactions.
If CP is violated, then all parameters which can be complex in the Lagrangian must be included.  For example, in a two-Higgs model, it is inconsistent
to have complex Yukawa couplings of the Higgs bosons to fermions and a real
Higgs-Higgs coupling $\mu^2 $. Fermionic loops will induce (infinite) complex
contributions to $\mu^2$ and these can only be absorbed if $\mu^2$ itself is
taken as a complex parameter.  So, renormalizability requires $\mu^2$ to be
complex.  

The upshot of these considerations is that, in general, the number of CP-violating phases entering into a theory increases with the complexity
of the theory.  In this respect, the three-generations CKM
model~\cite{CKM} is the simplest possible example of a CP-violating theory.
In the CKM model there are two CP-violating phases.  One of these is the CP-violating phase $\gamma$ (arising from the Yukawa interactions) which enters
in the quark mixing matrix $V_{\rm CKM}$.  The other phase is the CP-violating vacuum angle $\bar\theta$ which
accompanies the CP-odd gluon density $F\tilde F$.~\cite{tH}
Even in this most simple of models, however, we do not really understand
why these two CP-violating phases are so vastly different.  While $\gamma\sim
O(1)$, the parameter $\bar\theta < 10^{-10}$, so as not to to obtain too large
a dipole moment for the neutron.  This is the strong CP-problem.

If one ignores the strong CP problem, then it appears that the CKM
model gives a simple and consistent description of all existing
experimental data.  As we mentioned earlier, the CP-violating
phenomena in the neutral Kaon system are connected to the $\epsilon$ parameter,
which measures the CP even admixture in the $K_L$ state:
\begin{equation}
|K_L\rangle = \frac{1}{\sqrt{2}} (|K^o\rangle - |\bar K^o\rangle) +
\epsilon\frac{1}{\sqrt{2}}(|K^o\rangle + |\bar K^o\rangle) =
|K_2\rangle + \epsilon |K_1\rangle~.
\end{equation}
This parameter is experimentally small
$|\epsilon| \simeq |\eta_{00}|\simeq |\eta_{+-} |\sim 2 \times 10^{-3}$.~
\cite{PDG} In the CKM model this smallness is understood as resulting from the
smallness of the interfamily mixings, not because $\gamma$ is small. One finds:
$\epsilon\sim \lambda^4\sin\gamma\sim 10^{-3} \sin\gamma$,
where $\lambda$ is the sine of the Cabibbo mixing angle, $\lambda\sim 0.22$.

Nevertheless, the observed matter-antimatter asymmetry in the Universe
suggests that there are other significant CP violating phases, besides the
CKM phase $\gamma$.  As is well known,~\cite{Sakharov} to establish a non-zero
asymmetry $\eta_B$ one needs to have B- and CP-violating processes go out of 
equilibrium during the evolution of the Universe.  This can occur within the
framework of grand unified theories (GUTs), but could also have happened at
the time of the electroweak phase transition, if this transition was strongly
first-order.  In either case, it is easy to establish that $\eta_B$ necessarily
depends on other phases besides $\gamma$. This is clear in the case of GUTs, since these theories involve further
interactions beyond those of the standard model. For electroweak baryogenesis the argument is more subtle. It turns out that
if $\gamma$ is the only phase present at the electroweak phase transition then,
because of the GIM mechanism ~\cite{GIM}, $\eta_B$ is very small.  Typically, one finds~\cite{Shap}
\begin{equation}
\eta_B\sim\epsilon_{\rm CP~viol} \sim \lambda^6 \sin\gamma
\prod_{i<j} \frac{(m_j^2-m_i^2)}{(M_W)^{12}} \sim 10^{-18} \sin\gamma,
\end{equation}
so the generated asymmetry is negligible.  Furthermore, if there is only one
Higgs doublet, given the
present bound on $M_H \geq 90~{\rm GeV}$ from LEP200,~\cite{Higgs} the
electroweak phase transition is only weakly first order and even the tiny
asymmetry established gets erased.  Both these difficulties, in principle, can be obviated in multi-Higgs models.  However, these models introduce other CP-violating phases besides $\gamma$.

\section{CP-Violation---Testing the CKM Paradigm.}

The consistency of the CKM model with the observed CP-violating phenomena
in the Kaon system emerges from a careful study of constraints on the CKM
mixing matrix.  It is useful for these purposes, following Wolfenstein,\cite{Wolf} to expand the elements of $V_{\rm CKM}$ in powers of 
the Cabibbo angle $\lambda = \sin\theta_c = 0.22$:
\begin{eqnarray}
V_{\rm CKM}
&\simeq&
\left(
\begin{array}{ccc}
1-\lambda^2/2 & \lambda & A\lambda^3(\rho-i\eta) \\
-\lambda & 1-\lambda^2/2 & A\lambda^2 \\
A\lambda^3(1-\rho-i\eta) & -A\lambda^2 & 1
\end{array}
\right)
+ O(\lambda^4)~.
\end{eqnarray}
One sees from the above that, to $O(\lambda^4)$, the only complex phases
in $V_{\rm CKM}$ enter in the $V_{ub}$ and $V_{td}$ matrix elements:
\begin{equation}
V_{ub} = A\lambda^3(\rho-i\eta) \equiv |V_{ub}| e^{-i\gamma}; ~~~
V_{td} = A\lambda^3(1-\rho-i\eta) \equiv |V_{td}| e^{-i\beta}~.
\end{equation}
The unitarity condition
$\sum_i V^*_{ib}V_{id} = 0$
on the $V_{\rm CKM}$ matrix elements has a nice geometrical interpretation
in terms of a triangle in the $\rho-\eta$ plane with base $0\leq\rho\leq 1$
and with an apex subtending an angle $\alpha$, where $\alpha + \beta +
\gamma = \pi$.

One can use experimental information on $|\epsilon|$, the $B_d-\bar B_d$
mass difference, $\Delta m_d$, and the ratio of $|V_{ub}|/|V_{cb}|$ inferred
from $B$-decays to deduce a 95\% C. L. allowed region in the $\rho-\eta$ plane.
If one includes, additionally, information from the recently obtained
strong bound on $B_s-\bar B_s$ mixing [$\Delta m_s > 12.4~{\rm ps}^{-1} ~~$(95\% C.L.)~\cite{smix}], one further restricts the CKM allowed region.
Fig. 1 shows the result of a recent study for the Babar Physics Book.~\cite{Babar} As one can see, the data is consistent with a rather
large CKM phase $\gamma :~45^\circ \leq \gamma \leq 120^\circ~$.
If one were to imagine that $|\epsilon|$ is due to some other physics, as
in the superweak theory,~\cite{SW} then effectively the $\Delta S = 1$
parameter $\eta \simeq \gamma \simeq 0$. In this case one has another allowed region
for $\rho$ at the 95\% C.L.: $ 0.25 \leq \rho \leq 0.27$~.
\footnote{As we will discuss below, the non-zero result for $\epsilon^\prime/\epsilon$
recently announced by the KTeV Collaboration, ~\cite{KTeV} by itself excludes this superweak option.}

\begin{figure}
\begin{center}
\epsfig{file=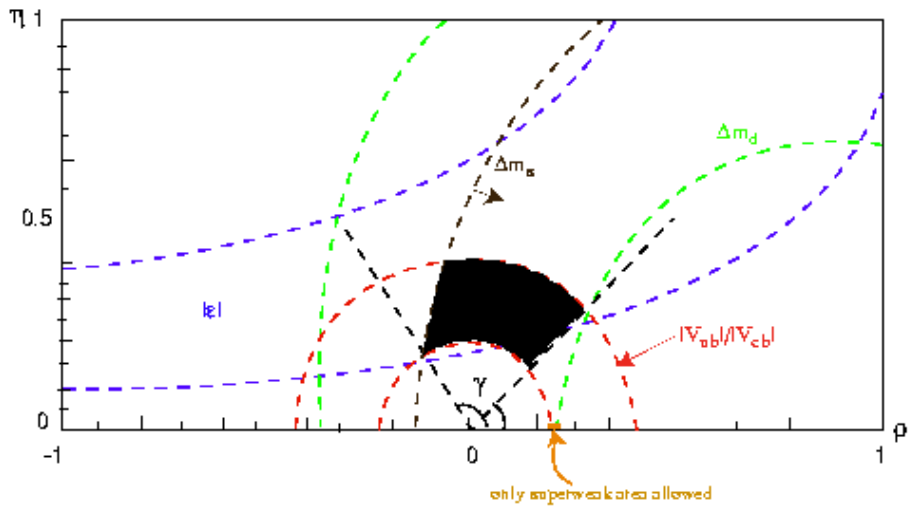,width=5in}
\end{center}
\caption[]{Allowed region in the $\rho-\eta$ plane. From Ref. [16]}
\end{figure}

The only CP violating constraint in Fig. 1 is provided by $|\epsilon|$.  So,
to really prove that the CKM phase $\gamma$ is large, one needs to measure
another CP-violating effect, which is independent of $|\epsilon|$.  In the
Kaon system this is provided by a measurement of $|\epsilon^\prime/\epsilon|$
and, eventually, of the extremely rare process $K_L\to \pi^0\nu\bar\nu$.
For B-decays, the simplest additional
CP-violating measurement involves extracting $\sin 2\beta$ from the study of
$B_d\to\Psi K_S$ decays.  I comment on each of these processes in what follows.

\subsection{$\epsilon^\prime/\epsilon$.}

At the time of WIN99 one expected news any day from the two large experiments
(KTeV and NA48) trying to measure $\epsilon^\prime/\epsilon$.  Both
experiments have run in the last two years and their aim is to
bring down the error for $\epsilon^\prime/\epsilon$ to 
a few times $10^{-4}$.  Hence, these experiments should be capable
to resolve the issue of whether $\eta$ is non-vanishing. 
Unfortunately, even with a meausurement of $\epsilon^\prime/\epsilon$ it will be difficult to get a good estimate for $\eta$. As is well known,~\cite{FR} $\epsilon^\prime/\epsilon$ depends
both on the contribution of gluonic Penguins and of electroweak Penguins. Although there is little uncertainty in the calculation of the coefficients of the relevant operators,  since all all attendant QCD effects have been computed to NLO, ~\cite{BB} the hadronic matrix elements of the Penguin operators
are not well determined. Furthermore, it turns out that these
two contributions tend to cancel, increasing the uncertainty of the
theoretical predictions.

It is useful to illustrate this with an approximate formula \footnote{For a more accurate treatment, see the review of Buchalla {\it et al.}~\cite{BB}} given some time ago by Buras and Lautenbacher.~\cite{BL}  These authors write for
Re $\epsilon^\prime/\epsilon$ the expression
\begin{equation}
{\rm Re}~\epsilon^\prime/\epsilon \simeq
[36\times 10^{-4}]A^2\eta\left\{B_6 - 0.175\left(\frac{m_t^2}{M_W^2}\right)^{0.93} B_8\right\}~.
\end{equation}
Here $B_6~(B_8)$ are the contribution of the gluonic
(electroweak) Penguins, appropriately normalized.  Using the CKM analysis of
Fig. 1, along with the measurement of $|V_{cb}|$ which fixes A, the parameter
$A^2\eta$ spans the range $0.18 \leq A^2\eta \leq 0.31$. Whence one predicts
\begin{equation}
{\rm Re} ~\epsilon^\prime/\epsilon \simeq
(6.5-11.3)\times 10^{-4} [B_6-0.75~B_8]~.
\end{equation}
In the vacuum insertion approximation $B_6=B_8=1$. Lattice calculations~\cite{C} and $1/N$ results~\cite{BB} give $B_6=B_8=1\pm 0.2$.
That is, they reproduce the vacuum insertion result to 20\%.  Unfortunately,
because of the cancellation between the gluonic and electromagnetic Penguin
operators, the corresponding uncertainty for $\epsilon^\prime/\epsilon$ in Eq.(6) is much greater. 

The experimental situation regarding $\epsilon^\prime/\epsilon$ was not
clear at the time of WIN99, since the old CERN ~\cite{NA31} and FNAL ~\cite{E731} results for
$\epsilon^\prime/\epsilon$ are mildly contradictory:
\begin{equation}
{\rm Re}~\epsilon^\prime/\epsilon =
\left\{
\begin{array}{ll}
(23 \pm 3.6 \pm 5.4) \times 10^{-4} & {\rm NA31}  \\
(7.4 \pm 5.2 \pm 2.9) \times 10^{-4} & {\rm E731} 
\end{array}
\right.
\end{equation}
However, naively, the theoretical formula (6) seems to favor the E731 result,
since it appears much easier to get Re $\epsilon^\prime/\epsilon\sim ~
{\rm few}~\times 10^{-4}$ than a much larger number.  Recently, however,
the KTeV Collaboration announced a new result for $\epsilon^\prime/\epsilon$
obtained from an analysis of about 20\% of the data collected in the last two
years.~\cite{KTeV}  Their result is
\begin{equation}
{\rm Re}~\epsilon^\prime/\epsilon = (28 \pm 3 \pm 2.6 \pm 1) \times
10^{-4} = (28 \pm 4.1)\times 10^{-4},
\end{equation}
where the first error is statistical, the second is systematic and the third
is an estimate of the error introduced by their Monte Carlo analysis.
Strangely, this result is much closer to the old CERN result than the
value obtained by E731, which was KTeV's precursor!  More importantly, the value obtained is nearly $7\sigma$ away from zero, giving strong
evidence that the CKM phase $\gamma$ is indeed non-vanishing.

Although it is not possible to extract a good value for $\eta$ from
this measurement, it is clear that the superweak solution in Fig. 1
is now excluded by this new data. However, the rather large value for
$\epsilon^\prime/\epsilon$ is theoretically surprising.  To get a value for Re $\epsilon^\prime/\epsilon$ as large as that of KTeV and NA48, either the cancellation between the gluonic and
electroweak Penguins is highly ineffective and/or the overall size of $B_6$ and
$B_8$ is much bigger than that suggested by the vacuum saturation
approximation.   Since $B_i \sim 1/m_s$ one can increase the overall size of these matrix elements by
supposing that the strange quark mass $m_s$ is much smaller than normally
assumed.  Whether this is warranted remains to be seen, although recent lattice estimates seem to point in this direction. ~\cite{K} 

\subsection{$K_L^0\to\pi^0\nu\bar\nu$.}

The (extremely) rare decay $K_L^0\to\pi^0\nu\bar\nu$ provides a much cleaner
measure of the CP-violating parameter $\eta$. 
Since $\pi^0J^*$, with $J^*$ a spin 1 virtual state, is CP odd, one can can write the amplitude for $K_L\to\pi^0J^*$ as~\cite{Dib}
\begin{equation}
A(K_L\to\pi^0J^*) = \epsilon A(K_1\to \pi^0J^*) + A(K_2\to\pi^0J^*),
\end{equation}
where the second term is non-vanishing only if there is direct CP
violation $(\eta\not= 0)$.  However, because of the factor of $\epsilon$ and
the smallness of the CP-conserving decay $K_1\to\pi^0\nu\bar\nu$, 
to a very good approximation, $A(K_L\to\pi^0\nu\bar\nu)\simeq A(K_2\to\pi^0\nu\bar\nu)$.
Furthermore, because the relevant $K$ to $\pi$ matrix element is well known
from the corresponding charged decay, there is little theoretical uncertainty.

The direct CP-violating amplitude is dominated by loops involving top.  Again, one can write an approximate formula for the branching ratio for this process, but now there is no matrix element uncertainty:  ~\cite{Buchbu}
\begin{equation}
B(K_L\to\pi^0\nu\bar\nu) = 8.07\times 10^{-11} A^4\eta^2
\left(\frac{m_t}{M_W}\right)^{2.3}~.
\end{equation}
A CKM analysis then leads to the prediction:
$1.5\times 10^{-11} < B(K_L\to\pi^0\nu\bar\nu) \leq 4.4\times 10^{-11}$.
Unfortunately, this theoretical expectation is about four orders of magnitude
below the present limit for this process, $B(K_L\to\pi^0\nu\bar\nu)
< 5.9\times 10^{-7}$. \cite{numbers} Given the difficulty of measuring this all neutral final state, it is difficult
to imagine trying to extract $\eta$ this way.  However, an experimental result
would yield a value for $\eta$ with very little theoretical error.

\subsection{$\sin 2\beta$.}

In my view, the measurement of $\sin 2\beta$ should
provide, in the near term, the strongest confirmation of the basic
correctness of the CKM paradigm.  First of all, the CKM analysis of Fig. 1
implies a rather large value for $\sin 2\beta$, since $\beta$ like $\gamma$
is a large angle.\footnote{For instance, a recent CKM analysis~\cite{smix}
gives $\sin 2\beta = 0.73 \pm 0.08$.}  Secondly, and of equal importance,
$\sin 2\beta$ is accessible experimentally in a theoretically clear context.
Let me briefly explain this last point.

As usual, to observe CP-violating affects one needs to have interference of
two different amplitudes with different CP phases.  This occurs naturally in
the time evolution of B-decays.  Consider specifically the decay of a state
$B_d^{\rm phys}(t)$ into some final state $f$.  The state $B_d^{\rm phys}(t)$
is defined by the property that at $t=0$ it was a $B_d$ state.  
Because  of mixing, $B_d^{\rm phys}(t)$ evolves in time as a linear superposition of $B_d$ and $\bar B_d$ states.  Thus the decays of $B_d^{\rm phys}(t)$ into the final state $f$ can follow two different paths, and their
interference can lead to CP-violating phenomena.

$B_d-\bar B_d$ mixing is dominated by the top quark box graph.  As a
result, the only large CP-violating phase which enters is the phase $\beta$
of $V_{td}$.  It is easy to show that, as a result of the mixing, one has
\begin{equation}
|B_d^{\rm phys}(t)\rangle = e^{-iM_Bt} e^{-\frac{\Gamma_B}{2} t}
\left\{\cos\frac{\Delta m_d}{2}t|B_d\rangle + i e^{-2i\beta}\sin
\frac{\Delta m_d}{2}t|\bar B_d\rangle\right\}
\end{equation}
It turns out that no other CP-violating phase will enter (in leading order in
$\lambda$) if the underlying processes $B_d\to f$ and $\bar B_d\to f$ only
involve the weak decays $b\to c\bar cs$ and $\bar b\to \bar c c\bar s$.  This
is obvious for the tree amplitudes, but it is also true for the 
$b\to s$ Penguin graph, since this is dominated by the top-loop and $V_{ts}$ and $V_{tb}$
are purely real.  In addition, for decays of
$B_d^{\rm phys}(t)$ to CP self-conjugate states $f$ ($\bar f = \eta_ff$, with
$\eta_f = \pm 1$) the $B_d$ and $\bar B_d$ amplitudes are simply related.  As
a result, one finds for these decays a formula with no theoretical 
ambiguities at all, namely
\begin{equation}
\left.\Gamma(B_d^{phys}(t)\to f)\right|_{b\to c\bar cs} =
\Gamma(B_d\to f)e^{-\Gamma_Bt}
\left\{1+\eta_f\sin 2\beta\sin\Delta m_dt\right\}~.
\end{equation}

There is great deal of interest in measuring $\sin 2\beta$ through the study of the
process $B_d^{\rm phys}(t)\to\psi K_S$, as this process has both a 
large branching ratio 
and a nice signature.  Of course, what is crucial for the
measurement is being able to correctly tag the decay as originating from an
initially produced $B_d$ or $\bar B_d$ state.  Last summer,
at the Vancouver Conference, both OPAL~\cite{OPAL} and CDF~\cite{CDF}
reported the first attempts at extracting $\sin 2\beta$.  Soon after WIN99, CDF announced the
result of an updated analysis of their data which, even with a
large error, is already quite interesting:
$\sin 2\beta = 0.79 \pm 0.44$.~\cite{CDF1}

A much more precise values for $\sin 2\beta$ should be
forthcoming soon from the B factories at KEK and SLAC.
In the B-factories one can measure $\sin 2\beta$ by looking at a variety of
modes besides $B_d\to\psi K_S$ (e.g. $B_d\to\psi K_L$; $B_d\to D\bar D$).  Of course, the precision with which each mode determines $\sin 2\beta$
differs, but consistency of the different results will
provide an important cross check.  A rather detailed analysis of the reach
achievable with the BABAR detector at the SLAC B Factory is contained in the
Babar Physics Book. ~\cite{Babar}  For instance, using just the $\psi K_S$ and $\psi K_L$ modes, one expects to be able to measure
$\sin 2\beta$ with an error $\delta\sin 2\beta = \pm 0.23$ ($\delta\sin 2\beta
= \pm 0.09$) with an integrated luminosity of 5 fb$^{-1}$ (30 fb$^{-1}$).
This accuracy is of the order of the present uncertainty in $\sin 2\beta$ from
a CKM analysis.  However, the B-factory results will be an {\bf actual mesurements} of CP violation!

\section{Successes and Challenges of Weak Decays.}

As we alluded to in the introduction, weak decays of heavy quark
systems are much more amenable to theoretical analysis.  In these systems,
a combination of a heavy quark expansion [HQET]~\cite{HQET} and perturbative QCD permits rather precise
predictions which, on the whole, have been well tested experimentally.  In
general, however, even here one has to restrict oneself to special theoretical
regions or inclusive enough processes. For example, predictions for exclusive decays, like the
process $B\to D^*\ell\bar\nu_\ell$, are valid only in the zero-recoil limit.
Similarly, predictions for inclusive decays, which rely on parton-hadron
duality, cease to be reliable near the kinematical limit. I will illustrate some of the theoretical issues with some examples.

\subsection{Extracting $|V_{cb}|$ from Inclusive Semileptonic Decays.}

Data on $B$ semileptonic decays coming from the
$\Upsilon(4s)$ and from $Z$ decays are mildly inconsistent.  As a result, the average width for $b\to c\ell\bar\nu_\ell$ decays
[$\Gamma(b\to c\ell\bar\nu_\ell) = (66.5\pm 3.0)~{\rm ps}^{-1}$~\cite{JRP}]
has still a 5\% error.  Since 
$\Gamma(b\to c\ell\bar\nu_\ell)\sim |V_{cb}|^2$ this experimental error 
implies a 2.5\% experimental error on $|V_{cb}|$. This error is comparable to the theory error quoted in different recent analyses. ~\cite{BBB,BSU}
If the data from CLEO and LEP could be reconciled,  potentially one should
be able to reduce the experimental error above by a factor of 2.  This raises the question of what accuracy one can hope to reach theoretically for $|V_{cb}|$?

The naive parton model formula for the B semileptonic width has a large uncertainty coming from the b-quark mass, since this enters raised to 
the $5^{\rm th}$ power, $\Gamma(b\to c\ell\bar\nu_\ell) \sim m_b^5$ .  This uncertainty is partially removed in the HQET. In this approach
$m_b$  is replaced by the B-quark mass $M_B$, and the parton model rate is modified by subleading $1/m_b$ corrections.  These latter
terms are proportional to matrix elements of local operators appearing
in the operator product expansion of the tensor $T_{\mu\nu}$:
\begin{equation}
T_{\mu\nu} = \int d^4x e^{iqx}
\langle B|\bar b(0)\gamma_\mu(1-\gamma_5)c(x)\bar c(0)\gamma^\mu(1-\gamma_5)
b(0)|B\rangle,
\end{equation}
whose imaginary part is related to the width $\Gamma$. ~\cite{HQET}  The upshot is that in HQET there are 2 new operators which enter the theory at
$O(1/m_b^2)$,~\footnote{One can show that there are no $O(1/m_b)$
corrections appearing in the HQET.} whose matrix elements one needs to know:
\begin{equation}
\lambda_1 = \frac{1}{2M_B} \langle B|\bar h_b(iD)^2
h_b|B\rangle ~;~~\lambda_2 = \frac{g_3}{12M_B} \langle B|\bar h_b \sigma^{\mu\nu} G_{\mu\nu} h_b|B\rangle.
\end{equation}
Here $D$ and $G_{\mu\nu}$ are the usual covariant derivative and field
strengths of QCD and $h_b(x) = (1+\gamma\cdot v)/2 e^{im_bv\cdot x} b(x)$, where
$v^\mu$ is the 4-velocity of the $b$-quark.
In addition, the relation between $m_b$ and
$M_b$ introduces a further parameter, $\bar\Lambda$, which, roughly speaking,
accounts for the momentum distribution of the $b$-quark in the $B$-meson.
One can write for the pseudoscalar and vector B-mesons the formulas ~\cite{HQET}
\begin{equation}
M_B =m_b+ \bar\Lambda - \frac{\lambda_1+\lambda_2}{2m_b} ~;~~
M_B^* = m_b + \bar\Lambda - \frac{\lambda_1-\lambda_2}{2m_b}
\end{equation}
The hyperfine splitting between
$M_B^*$ and $M_B$ serves to fix $\lambda_2$ and one finds $\lambda_2\simeq 0.12~{\rm GeV}^2$. So, at this level of accuracy, there are ony two free parameters left, $\lambda_1$ and $\bar\Lambda$.

Provided one has estimates for $\lambda_1$ and $\bar\Lambda$, the formula for the semi-leptonic width (including $O(1/m^2_b)$ terms and QCD corrections to $O(\alpha_s)^2$) ~\cite{CM} can be used
to determine $|V_{cb}|$. One can extract values for these parameters by studying
the moments of the hadronic mass spectrum ~\cite{Falk}
\begin{equation}
\langle (s_H-\bar M_B^2)^n\rangle =
\frac{1}{\Gamma(B\to X_c\ell\bar\nu_\ell)} \int ds_H
\frac{d\Gamma}{ds_H} (s_H-\bar M_B^2)^n~,
\end{equation}
where $s_H$ is the invariant mass of the recoiling hadronic state and
$\bar M_B^2 = \frac{1}{4}(M_B^2+3M_B^2)$.
Similarly, moments of the electron energy spectrum ~\cite{Vol} also depend on  
$\lambda_1$ and $\bar\Lambda$.

Recently, the CLEO Collaboration ~\cite{Poling} has meausured the
first and second moments of the hadronic mass spectrum, obtaining for $\lambda_1$
and $\bar\Lambda$, the following values:
\begin{equation}
\bar\Lambda = (0.33 \pm 0.02 \pm 0.08)~{\rm GeV}~; ~~~
\lambda_1 = -(0.13 \pm 0.01 \pm 0.06)~{\rm GeV}^2~.
\end{equation}
These results are similar to those obtained by  Gremm {\it et al.}~\cite{GKLN}
by considering moments of the lepton spectrum integrated over a restricted energy range: [$ \bar\Lambda = (0.39 \pm 0.11)~{\rm GeV}~;~
\lambda_1 = -(0.19 \pm 0.10)~{\rm GeV}^2~]$. Unfortunately,
the situation is a bit confusing at the moment, since CLEO obtained a different set of values  $[\bar\Lambda\simeq 1~{\rm GeV};~
\lambda_1\simeq -0.8~{\rm GeV}^2$~\cite{Poling}] by
considering leptonic moments where one integrates over (nearly) all the lepton
energy distribution.  These preliminary results, however, may be an
artifact since they are quite sensitive to
corrections coming from the unmeasured pieces of the leptonic spectrum.

If one just uses the values of $\bar\Lambda$
and $\lambda_1$ from the hadronic energy spectrum analysis, then from the experimental
value for the width $\Gamma(b\to c\ell\bar\nu_\ell)$, one deduces ~\cite{Ligeti}
that $|V_{cb}| = 0.0415 \pm 0.0010 \pm 0.0010$.
So, indeed, the theoretical and experimental errors on $|V_{cb}|$ are
comparable. Nevertheless, one should note~\cite{Ligeti} that if indeed
$\bar\Lambda \simeq 1$ GeV and $\lambda_1\simeq -0.8~{\rm GeV}^2$, then
the value of $|V_{cb}|$ goes up by 7\%! However, because the above value is quite  consistent with that extracted from an exclusive analysis of the decay
$B\to D^*\ell\bar\nu_\ell$ at zero recoil [$|V_{cb}| = 0.0387 \pm 0.0031$~\cite{Drell}],
this suggests that the CLEO full lepton moment analysis is probably flawed.

\subsection{Extracting $V_{ub}$ from $B$-Decays.}

Because the decay $b\to ul\bar\nu_\ell$ is highly suppressed relative to $b\to c\ell\bar\nu_\ell$, to extract $|V_{ub}|$ from exclusive decays one must focus
on the limited kinematical region where the decays to charmed states are forbidden.
This restricted region is characterized by having the lepton energy near its
upper end point $[E_\ell > (M_B^2-M_D^2)/2M_D]$ and the produced hadronic mass
squared $s_H < M_D^2$.  Falk, Ligeti and Wise, ~\cite{FLW}  have suggested
that concentrating on the second kinematical restriction is better than just looking at the high energy end of the lepton spectrum.  This is because for $s_H < M_D^2$ a continuum of states contribute and hadron-parton duality should be reliable.
On the other hand, at the end of the lepton spectrum typically $\pi$ and
$\rho$ exclusive states dominate.

The differential rate for B-decays into final states of a given hadronic mass
squared $s_H$, for $\bar\Lambda M_B< s_H < M_D^2$, can be expressed in terms of a
shape function $S(s_H,\bar\Lambda)$ which is universal in character and
depends on the $b$-quark momentum distribution parameter $\bar\Lambda$:~\cite{FLW}
\begin{equation}
\frac{d\Gamma}{ds_H} = \frac{G_F^2M_B^3}{192\pi^3} |V_{ub}|^2
\left(1-\frac{\bar\Lambda}{M_B}\right)^3 S[s_H,\bar\Lambda]~.
\end{equation}
If indeed $\bar\Lambda\sim 300-400~{\rm MeV}$ and its
error can be kept to the level of $\delta\bar\Lambda \leq 50~{\rm MeV}$,
then it should be possible ~\cite{Ligeti} to reduce the error on $|V_{ub}|$
to around 10\%.  This would be a considerable improvement over the
present day exclusive determination of this matrix element from the decays
$B\to\pi\ell\bar\nu_\ell,~B\to\rho\ell\bar\nu_\ell$ [$|V_{ub}| = (3.3 \pm 0.2 \textstyle^{+0.3}_{-0.4} \pm 0.7) \times 10^{-4}$~\cite{Drell}]
which contains a 20\% model error.

\subsection{$B\to X_s\gamma$.}

Last year saw a refined measurement from CLEO~\cite{Alexander} (as well a
first ALEPH result~\cite{ALEPH}) on this important process.  The 
branching ratio obtained $[{\rm BR}(B\to X_s\gamma) = (3.15 \pm 0.35 \pm 0.32 \pm 0.26) \times 10^{-4}]$ includes an estimate of the error introduced by having
to extrapolate to below $E_\gamma = 2.1$ GeV, where one cuts on the data.  At
the same time, a host of theoretical refinements became available bringing
the theoretical expectations for $B\to X_s\gamma$ also into excellent
shape.  Thus a meaningful and stringent comparison between theory and
experiment is now possible.

The QCD corrections to the effective Lagrangian describing this process
\begin{equation}
{\cal{L}}_{\rm eff} = \frac{4G_F}{\sqrt{2}} V_{ts} V^*_{tb}c_7^{\rm eff}
(\mu) \frac{e}{16\pi^2} \bar s\sigma^{\nu\mu} m_s
\frac{(1+\gamma_5)}{2}b F_{\mu\nu}
\end{equation}
are very important, but they are controllable and known.  For example,
the coefficient $c_7^{\rm eff}(\mu)\simeq -0.19$, without taking into account
of QCD.  This number changes by more than 30\% when lowest order QCD effects are
incorporated, and is quite $\mu$-dependent.  Fortunately, the full NLO QCD
effects have now been calculated by a number of groups~\cite{bsgamma} leading
to a stable result for $c_7^{\rm eff}(\mu)$, with little $\mu$-dependence.  As a
result, theoretically, the branching ratio for $B\to X_s\gamma$ is now known
with an error of 10\%---comparable to that of the CLEO result
$[{\rm BR}\left.(B\to X_s\gamma)\right|_{\rm theory} =
(3.29 \pm 0.34)\times 10^{-4}$ ~\cite{Neubert}].
Since this result is in excelent agreement with experiment, there is little room
for beyond the Standard Model contributions.

The theoretical result for the $B\to X_s\gamma$ branching ratio
is not that sensitive to $1/m_b^2$ corrections which, typically, are of
$O(5\%)$.~\cite{FLS}  However, as Kagan and Neubert~\cite{KN} have pointed out, the
differential branching ratio as a function of the photon energy is quite sensitive
to the momentum distribution of the $b$ quark in the $B$-meson.  As a result, the photon energy spectrum in $B\to X_s\gamma$ can give information
on the nonperturbative parameters $\bar\Lambda$ and $\lambda_1$.  This is apparent from the recent analysis of
Neubert ~\cite{Neubert} which finds that the errors for the branching ratio for
$B\to X_s\gamma$ due to $O(1/m_b^2)$ effects is about 10\% if the cut on
$(E_\gamma)_{min}$ is at 2.2 GeV, but is less than 5\% when that cut is
reduced to 2.0 GeV.  More interestingly,~\cite{Neubert,Bauer} one could imagine
using the shape function obtained from analyzing the photon energy spectrum in
$B\to X_s\gamma$ to extract from
the differential rate for $B\to X_u \ell\bar\nu_\ell$ a more accurate value of
$|V_{ub}|$. 

\section{Looking for the New CP-Violating Phases.}

If the CKM model is correct, one expects the unitarity triangle to close, so that
$\alpha + \beta + \gamma = \pi$.  However, if there are other CP-violating
phases arising from new physics,
one can expect to alter this simple relation between the angles in the
unitarity triangle.  As we discussed earlier, the strongest predictions which
emerge from a CKM analysis of the present data is that both the angles $\beta$ and
$\gamma$ are rather large.  Even so, it might well be that when one measures
$\sin 2\beta$ with accuracy it will not agree with the value of $\sin 2\beta$
coming from the CKM analysis.  In what follows, I want to briefly discuss how
this might come about.

Recall from Eq. (11) that the phase $2\beta$ entered in the time evolution
of $B_d^{\rm phys}(t)$ as the CKM phase connected with $B_d-\bar B_d$ mixing.  It is possible that other physics enters in this mixing beyond the CKM
model, bringing additional CP-violation phases.  Let
us call the additional CP-violating phase entering in $B_d-\bar B_d$ 
mixing $\theta_M$.  Then, effectively, everywhere one should replace
$\sin 2\beta\to\sin 2(\beta + \theta_M)$.  Unfortunately, even pushing
parameters to extremes, it is difficult to generate a very large ``new physics"
CP-violating phase $\theta_M$.  For instance, in supersymmetric models at
most $\theta_M\sim 10^\circ$.~\cite{susy}  Since the CKM $\beta$-range is
$10^\circ \leq \beta\leq 27^\circ$, even such a large phase $\theta_M$ is
difficult to detect! \footnote{Note that since the extra mixing phase
$\theta_M$ obviously does not affect the CKM angle $\gamma$, one expects also
that $\sin 2\alpha\to\sin 2(\alpha-\theta_M)$,  However, $\alpha$ is even more
uncertain than $\beta$ and these effects are even harder to pin down.}

More promising than the new mixing phases $\theta_M$, are phases arising from new
physics which may affect Penguin amplitudes.  A good example is provided by
the pure Penguin decay $b\to s\bar ss$.  In the CKM model, the $b\to s$ Penguin
amplitude is dominated by top loops and is purely real.  However, in
supersymmetry $\tilde b-\tilde s$ mixing can bring additional CP-violating
phases and it is possible that the amplitude ratio $ A(B_d\to\phi K_s)/A(\bar B_d\to\phi K_s)\equiv e^{i\Phi_p}$
reflect this ``new" Penguin phase $\Phi_p$.  Because Penguin effects are
subdominant in processes like $B_d\to\psi K_S$, the phase $\Phi_p$ is
probably unimportant in this process.  Hence the time evolution of
$B_d^{\rm phys}(t)\to\psi K_S$ essentially still measures $\sin 2\beta$.
However, the time evolution of $B_d^{\rm phys}(t)\to\phi K_S$ actually
measures $\sin(2\beta + \Phi_p)$.  So one can look for new physics
CP-violating phases by comparing the values of the coefficient of $\sin\Delta m_dt$ in these two processes.  Obviously, error control is crucial.  

There are many other examples of such strategies.  For instance, in the
CKM model, the time evolution of $B_s^{\rm phys}(t)$ into $\psi\phi$ should
show no $\sin\Delta m_dt$ term, since there is no mixing phase for
$B_s^{\rm phys}(t)$ and there are no decay phases for $b\to c\bar cs$.  Finding
such a term could provide evidence for new physics.
This discussion raises the issue of how well one can test the 
unitarity triangle by measuring directly, in addition to $\beta$, also
$\alpha$ and $\gamma$.  Let me consider each of these angles in turn.

\subsection{$\alpha$.}

In principle, the angle $\alpha$ is measurable in an analogous way to
$\beta$.  One now needs to study the decays of $B_d^{\rm phys}(t)$ into
final states that can be accessed through a $b\to u$ transition.  A good
example is provided by $B_d\to\pi^+\pi^-$.  If the quark decay
$b\to u\bar ud$ is dominated by the tree amplitude, then
\begin{equation}
\frac{A(\bar B_d\to \pi^+\pi^-)}{A(B_d\to \pi^+\pi^-)} =
\left.\frac{A(b\to u\bar ud)}{A(\bar b\to\bar uu\bar d)}\right|_{\rm Tree} =
\frac{V_{ub}}{V_{ub}^*} = e^{-2i\gamma}~.
\end{equation}
This extra decay phase $e^{-2i\gamma}$ adds to the contribution from the
$B_d-\bar B_d$ mixing phase $e^{-2i\beta}$.  Using that
$\alpha + \beta+\gamma=\pi$, one finds
\begin{equation}
\left.\Gamma(B_d^{\rm phys}(t)\to\pi^+\pi^-)\right|_{\rm Tree} =
\Gamma(B_d\to\pi^+\pi^-)e^{-\Gamma_Bt}
\{1-\sin2\alpha\sin\Delta m_dt\}~.
\end{equation}

For $B_d\to\pi^+\pi^-$, however, one cannot neglect the effects of
the Penguin graphs, since for  $b\to u\bar ud$ decays these graphs have a different phase structure than the
tree graphs.~\cite{GLP}  While the tree graph phase is that of $V_{ub}$,
$e^{-i\gamma}$, the $b\to d$ Penguin has a phase $e^{i\beta}$---the phase of
$V_{td}^*$ entering in the dominant $t$-quark loop.  Because these two
phases are different, it is important to try to understand the effects of
this ``Penguin pollution".
Penguin pollution will alter Eq. (21) in two ways.  Consider the parameter
$\xi = e^{-2i\beta} A(b\to u\bar ud)/A(\bar b\to\bar uu\bar d)$.  Ignoring
Penguins, $\xi$ is simply $e^{-2i(\beta + \gamma)} = e^{2i\alpha}$.  Including
Penguins, $\xi$ becomes
\begin{equation}
\xi = e^{2i\alpha}[\frac{1+\frac{|P|}{|T|} e^{-i\alpha}e^{i\delta}}
{1+\frac{|P|}{|T|} e^{i\alpha}e^{i\delta}}] =
|\xi| e^{2i\alpha_{\rm eff}}
\end{equation}
where $\delta$ is an (unknown) strong interaction phase.  Because
$|\xi|$ is not unity now, and $\alpha\not= \alpha_{\rm eff}$, the rate for
$B_d^{\rm phys}(t)\to\pi^+\pi^-$ will now also have a $\cos\Delta m_dt$
term, as well as a modified $\sin\Delta m_dt$ term.  It is easy to see that the
rate formula becomes
\begin{equation}
\Gamma(B_d^{\rm phys}(t)\to\pi^+\pi^-) =
\Gamma(B_d\to\pi^+\pi^-) e^{-\Gamma_Bt}
\{1+a_c\cos\Delta m_dt-a_s\sin\Delta m_dt\}
\end{equation}
where
\begin{equation}
a_c = \frac{1-|\xi|^2}{1+|\xi|^2}~; ~~~~
a_s = \frac{2|\xi|}{1+|\xi|^2} \sin 2\alpha_{\rm eff}~.
\end{equation}

Gronau and London~\cite{GL} suggested estimating Penguin pollution
in the $B_d\to\pi\pi$ process through an isospin analysis. Their idea  is simple to describe.
If one neglects electroweak Penguins, then isospin is a good quantum number and one
can use isospin to classify the various decay amplitudes. 
$A(B^+\to\pi^+\pi^0)$ is a $\Delta I=3/2$ amplitude, and as such must be
purely given by tree graphs, since the Penguin graphs are $\Delta I=1/2$.  
Because the phase of the tree graphs is that of $V_{ub}$, it follows that
$A(B^+\to\pi^+\pi^0) = e^{2i\gamma}
A(B^-\to\pi^-\pi^0) $. 
Isospin, in addition, relates the decay modes of $B_d$, $B_{\bar d}$ into $\pi\pi$ to the charged $B$ decays:
\begin{eqnarray}
\frac{1}{\sqrt{2}} A(B_d\to\pi^+\pi^-) &+& A(B_d\to\pi^0\pi^0)
= A(B^+\to\pi^+\pi^0) \nonumber \\
\frac{1}{\sqrt{2}} A(\bar B_d\to\pi^+\pi^-) &+&
A(\bar B_d\to\pi^0\pi^0) = A(B^-\to\pi^-\pi^0)~.
\end{eqnarray}
These expressions geometrically can be represented as two triangles in
the complex plane, with a common base. It is easy to check that the misallignment angle between these triangles is related to the
phase $\alpha_{\rm eff}$. ~\cite{GL}  Hence, from measurements of all the relevant rates one can infer the Penguin pollution.

There have been a number of model studies to see what kind of errors one
might expect on $\alpha$.  One of the most complete of these studies is that
done in the Babar Physics Book,~\cite{Babar} where a variety of decay modes
$[B_d\to\pi\pi, \rho\pi, \rho\rho, a_1\pi]$ were considered.  Because the
relevant branching ratios are not known, some reasonable assumptions had to be
made both for these quantities and to estimate the amount of Penguin
pollution.  Assuming an integrated luminosity of 30 ${\rm fb}^{-1}$ the
resulting typical error expected for $\delta a_s$ for the $\pi^+\pi^-$
mode was $\delta a_s\simeq 0.26$, while for the $\rho^0\rho^0$ mode this
error was $\delta a_s \simeq 0.17$.  Unfortunately, it is difficult to extrapolate from these
results the expected error on $\sin 2\alpha$ since the connection of
$\delta\sin 2\alpha$ to $\delta a_s$ is itself channel- (and model-) 
dependent.  Nevertheless, it appears difficult to imagine measuring
$\sin 2\alpha$ to better than $\delta\sin 2\alpha = 0.2$.

\subsection{$\gamma$.}

The situation with the angle $\gamma$ is perhaps even more challenging, but
at the same time more interesting.  A number of authors have suggested trying
to extract $\gamma$ by looking at various asymmetries in processes which are
dominated by tree amplitudes, but where the final state is not CP self-conjugate.  One suggestion~\cite{GLAK} is to
study the time evolution of $B_s^{\rm phys}(t)\to D\phi$, in which both
the $\bar b\to\bar uc\bar s$ and $b\to c\bar us$ processes participate.
This process is sensitive to $\sin\gamma$, but it is very challenging
experimentally both because it involves $B_s$ mesons and because of the very
rapid $B_s-\bar B_s$ oscillations.  Bigi and Sanda, as well as Sachs, ~\cite{SBS} suggested instead studying the time evolution of $B_d^{\rm phys}(t)$ into
$D^{*\pm}\pi^\mp$.  Here the processes $\bar b\to\bar uc\bar d$ and
$b\to c\bar ud$ are involved and the time evolution measures
$\sin(2\beta+\gamma)$.  In this later example, the experimental challenge is
that the predicted effect is very small.

Alternatively, as suggested by Gronau and Wyler,~\cite{GW} one can try to
extract $\gamma$ by using triangle relations involving charged $B$-decays,
similar to those we discussed earlier for $B_d\to\pi\pi$.  This is
nicely illustrated by the processes $B\to DK$, although the effects involved
are probably not measurable experimentally.  Both the decays
$B^+\to\bar D^0K^+$ and $B^-\to D^0K^-$ are pure tree decays, involving
$\bar b\to\bar cu\bar s$ and its complex conjugate.  Since $V_{cb}$ is real,
it follows that $A(B^+\to\bar D^0K^+)=A(B^-\to D^0K^-)=A_1$, where $A_1$
can be taken as real by convention.  On the other hand, the decays
$B^+\to D^0K^+$ and $B^-\to\bar D^0K^-$, which are governed by the tree
process involving $\bar b\to\bar uc\bar s$ and its complex conjugate, involve
$V_{ub}$ and hence the phase $\gamma$.  Hence one has
\begin{equation}
A(B^+\to D^0K^+) = A_2 e^{i\gamma}e^{i\delta} ~;~~A(B^-\to\bar D^0K^-) =
 A_2 e^{-i\gamma}e^{i\delta}~,
\end{equation}
where $\delta$ is a strong rescattering phase.  It is easy to see that by
measuring the rates for two of the above processes, as well as the rates for
$B^+\to D^0_+K^\pm$, where $D_+^0 = \frac{1}{\sqrt{2}}(D^0+\bar D^0)$ is
a CP eigenstate, one can reconstruct $\gamma$ essentially by trigonometry.~\cite{GW}  Unfortunately, this will not work in practice because
the triangles are too squashed.  Furthermore,~\cite{ADS} these decays are
affected by rescattering effects which further complicate matters.

One can apply these ideas to channels with bigger branching ratios.
However, in general, now one has both tree and Penguin contributions.  Perhaps one of the nicest
examples is provided by $B\to \pi K$, where lots of interesting dynamical
features appear.~\cite{pik} I want to illustrate some of the issues involved in these decays by discussing the,
so called, Fleischer-Mannel bound~\cite{FM} on $\sin^2\gamma$.  If one retains
only the gluonic Penguins and neglects altogether rescattering effects, then
one has simple formulas for the decays $B^+\to\pi^+ K^0;B_d\to \pi^-K^+$
and $\bar B_d\to\pi^+K^-$.  The first decay is purely a Penguin process and,
because the $\bar b\to\bar s$ Penguin is dominated by the top quark, there
is no CP phase.  The other two decays involve both trees and Penguins, with
the tree amplitude having the phase of $V_{ub}^*$ or $V_{ub}$, respectively.
Thus one can write $A(B^+\to\pi^+K^0) = P $, while $A(B_d\to\pi^-K^+) = -[P+Te^{i\gamma}e^{i\delta}]$ and
$A(\bar B_d\to\pi^+K^-) = -[P+Te^{-i\gamma}e^{i\delta}]$, 
where $\delta$ is an (uncalculable) strong rescattering phase between the Penguin and tree contributions.

Using the above, the Fleischer-Mannel ratio $R$ is easily computed.  One
finds
\begin{equation}
R = \frac{\Gamma(B_d\to\pi^-K^+) + \Gamma(\bar B_d\to\pi^+K^-)}
{\Gamma(B^+\to\pi^+K^0) + \Gamma(B^-\to\pi^-\bar K^0)}
= 1 + 2r\cos\delta\cos\gamma + r^2
\end{equation}
where $r = T/P$.  If $R < 1$, there is clearly negative interference
between the Penguin and tree amplitudes and one can get a bound on $\gamma$.
Indeed, it is easy to show that this bound is:  $R\geq\sin^2\gamma$.~\cite{FM}  Present day data from 
CLEO~\cite{FW} is tantalyzing since it gives
$R = 1.0 \pm 0.4 \pm 0.2 \pm 0.1$.
However, even if the data were to get more precise, matters are not as simple  because $R$ receives important corrections both from electroweak
Penguins and from rescattering effects.~\cite{Recorr}  

Rescattering in the $\pi K$ system can change $\pi^0K^+$ into
$\pi^+K^0$.  This, effectively, leads to the replacement of
the Penguin amplitude P by~\cite{N2}
$\tilde P[1+\epsilon_ae^{i\gamma} e^{i\delta_a}]$.
Here the parameter $\epsilon_a$ (and the strong interaction phase $\delta_a$) are
a measure of the rescattering and $e^{i\gamma}$ is the phase of
$V_{ub}^*$.  Electroweak Penguins have no weak CP phase, but introduce an additional strong
interaction phase.  Effectively they can be accounted for by the replacement:
$Te^{\pm i\gamma}e^{i\delta}\to Te^{i\delta}
[e^{\pm i\gamma} + q_{EW}e^{i\delta_{EW}}]$.
Here $\delta_{EW}$ is another strong interaction phase, while $q_{EW}$ parametrizes
the strength of these contributions.  These changes alter the Fleischer-Mannel bound to:
$R\geq F(\epsilon_a;q_{EW};\delta_a;\delta_{EW}) \sin^2\gamma$,
where F is a calculable function of these new parameters.  Neubert~\cite{N2}
has argued that the rescattering effects are small $(\epsilon_a\leq 0.1)$,
but that $q_{EW}$, in fact, can be large $(q_{EW} \sim 0.5)$. If this is so,
the Fleischer-Mannel bound is significantly affected.  However, a somewhat different
ratio studied by Neubert and Rosner~\cite{NR} appears to be more robust.

Given the uncertainties in all the methods discussed, it is clear that it is
difficult to estimate the accuracy one may ultimately obtain for $\gamma$.
Nevertheless, because information on this angle can be obtained in a variety of ways, this may help narrow down a range for the CKM phase $\gamma$. Nevertheless,
I remark that the extensive discussion presented in the Babar Physics
Book~\cite{Babar} on $\gamma$ only ended up by hazarding a guess on the accuracy which might be achieved. It is suggested there that, with
lots of integrated luminosity (100 fb$^{-1}$), perhaps one could hope to
determine $\gamma$ to $\delta\gamma = \pm (10-20)^\circ$.

\section{Concluding Remarks.}

It is clear that much theoretical progress has been made to control uncertainties in the predictions for weak decays and CP-violation.  Particularly for
the $B$-system a combination of beyond the leading order QCD corrections and
HQET, in specific and controlled circumstances, can give results with rather
small theoretical errors.  These results, in turn, allow for the extraction
from the data of fundamental parameters, like the elements of the CKM
matrix.

This said, however, one has to admit that our theoretical understanding of
CP-violation is still very rudimentary.  We have no real explanation of why
$\bar\theta < 10^{-10}$, unless axions are really found; we also have no real
clue if there are any other low-energy CP-violating phases besides the CKM
phase $\gamma$---and even for $\gamma$ our evidence is still rather tentative.
Fortunately, we are at the threshold of a new era of experimentation.  As we discussed, very
recently KTeV announced a value for $\epsilon^\prime/\epsilon$ and this should
be followed shortly by a similar announcement from NA48.  Furthermore, the
Frascati $\Phi$ Factory and its detector KLOE should soon be producing data.
On the $B$-decay front, CLEO keeps integrating luminosity and adding to our
detailed knowledge of these decays.  At the same time, very soon both the SLAC
and KEK $B$-factories, with their detectors BABAR and BELLE, should be
running providing new information on $B$ CP-violation.  The remarkable recent
result on $\sin 2\beta$ from CDF argues that also the Tevatron, in its
forthcoming run with the  Main Injector, will contribute substantially in
this area.  So there is real hope that experiment will shed some clarifying
light soon on the nature of CP-violation.  Let us hope so!

\section*{Acknowledgments.}

This work was supported in part by the Department of Energy under Contract
No. DE-FG03-91ER40662, Task C.

\end{document}